\documentclass[a4paper, 10pt, twocolumn]{article}

\usepackage{amsfonts,amssymb,amsmath,amsthm}
\usepackage{subfigure}
\usepackage[footnotesize]{caption}
\usepackage{cite}

\usepackage{pgfplots,pgfkeys,tikz,graphicx}
\usepackage{filecontents}
\usepackage{xcolor,url}
\usepackage{algorithm,algorithmic} 

\newcommand{\normF}[1]{\left|\left| #1 \right|\right|_\mathcal{F}}

\newcommand{\R}{\mathbb{R}}

    \newenvironment{customlegend}[1][]{%
        \begingroup
        \csname pgfplots@init@cleared@structures\endcsname
        \pgfplotsset{#1}%
    }{%
        \csname pgfplots@createlegend\endcsname
        \endgroup
    }%

    \def\addlegendimage{\csname pgfplots@addlegendimage\endcsname}
\pgfplotsset{
cycle list={%
{draw=brown, solid,line width=2pt}, 
{draw=brown, dashed,line width=2pt},
{draw=black,solid,line width=2pt},
{draw=black, dashed,line width=2pt},
{draw=blue,solid,line width=2pt},
{draw=blue, dashed,line width=2pt},
{draw=red,solid,line width=2pt},
{draw=red, dashed,line width=2pt},
}}

\usepackage[top=1.5cm, left=1.5cm, right=1.5cm, bottom=1.5cm]{geometry}

\title{Gaussian Compression Stream: Principle and Preliminary Results}

\author{Farouk Yahaya, Matthieu Puigt, Gilles Delmaire, and Gilles Roussel\\
\footnotesize Univ. Littoral C\^ote d'Opale, LISIC -- EA 4491, F-62228 Calais, France
}
\date{\empty} 

\renewenvironment{abstract}{\bf\small {\em\ Abstract---}}{}

\begin{document}

\maketitle

\begin{abstract} Random projections became popular tools to process big data. In particular, when applied to Nonnegative Matrix Factorization (NMF), it was shown  that structured random projections were far more efficient than classical strategies based on Gaussian compression. However, they remain costly and might not fully benefit from recent fast random projection techniques. In this paper, we thus investigate an alternative to structured random projections---named \textit{Gaussian compression stream}---which (i) is based on Gaussian compressions only, (ii) can benefit from the above fast techniques, and (iii) is shown to be well-suited to NMF.
\end{abstract}

\section{Introduction}
\label{sec:introduction}

Dimension reduction techniques are the linchpin for solving problems involving high dimensional data. They can capture most of the important features of the underlying high data while providing the benefit of mapping onto a much lower dimensional space, due to their computational intricacies and geometric properties. Among the numerous techniques proposed in the literature, those based on randomized linear algebra \cite{Halko_2011,Drineas_2016} have been shown to be particularly efficient. 

In particular, they were successfully combined to Nonnegative Matrix Factorization (NMF) \cite{Tepper_2016,Erichson_2018,Yahaya_2018} and its weighted extension \cite{Yahaya_2019} and is sometimes named \textit{compressed NMF}. More precisely, it was shown in \cite{Tepper_2016} that \textit{structured} random compression---based on Randomized Power Iteration (RPI)---was far more efficient than classical Gaussian Compression (GC) when applied to NMF. However, while several strategies have been proposed to speed-up GC---\textit{e.g.}, CountGauss \cite{Kapralov_2016} or specific hardware \cite{Saade_2016}---RPIs still suffer from a high computational cost. This analysis remains true when considering a stable extension of RPIs named Randomized Subspace Iterations (RSIs) \cite{Halko_2011}. 
In this paper, we 
propose an extension of GC nammed GC Stream (GCS). We show that GCS allows a NMF performance which is similar to RPIs/RSIs. We lastly discuss conditions where GCS should outperform RPIs/RSIs when the computational complexity is taken into consideration. 

\section{Compressed NMF}
\label{sec:first-section}

NMF is a popular signal \& image processing / machine learning tool which consists of estimating two $n \times p$ and $p \times m$ nonnegative matrices $G$ and $F$, respectively, from a $n\times m$ nonnegative matrix $X$ such that $X \approx G \cdot F$ \cite{Wang_2013,Gillis_2014}. %
While several cost functions and additive constraints have been proposed 
to that end, in its basic form involving the Frobenius norm $\normF{\cdot}$, NMF usually consists of solving alternating subproblems, \textit{i.e.}, 
\begin{align}
\hat{G} & =  \arg\min_{G \geq 0} \normF{X - G \cdot F}, \label{eq:NMF_update1} \\
\hat{F} & =  \arg\min_{F \geq 0} \normF{X - G \cdot F}, \label{eq:NMF_update2}
\end{align}
When $X$ is large, several strategies have been proposed to speed-up the updates, \textit{e.g.}, distributed \cite{Liu_2010} or online \cite{Mairal_2010} computations, fast solvers \cite{Guan_2012}, or randomized 
techniques \cite{Tepper_2016}. 

Actually, several randomized strategies were proposed in the literature. In \cite{Zhou_2012}, the authors assumed that $X$ is low-rank and can be replaced by a product $A \cdot B$ which helps the NMF factors to be cheaper to update, and which can be efficiently computed using randomized SVD. In \cite{Wang_2010}, the authors introduced the concept of \textit{dual random compression} described in Algorithm~\ref{algo:compressed_nmf}. The key idea consists in noticing that compressing $X$ by a projection on the left or the right side still allows to estimate the full matrix $F$ or $G$, respectively. The difficulty then lies in designing efficient matrices $L$ and $R$: the authors in \cite{Wang_2010} used scaled Gaussian realizations as tentative compression matrices, thus following the general proof of the Johson-Lidenstrauss Lemma (JLL) \cite{Johnson_1984} on which is built the theory of random projections. The authors in \cite{Tepper_2016,Erichson_2018} then found that adding some structure on the compression matrices allows a much better NMF performance (with different tested solvers). To that end, they used RPIs \cite{Halko_2011}. To compute $L$, RPIs are defined as 
\begin{equation}
L \triangleq \text{QR}\left( (X X^T)^q \cdot X \cdot \Omega_L \right)^T,\label{eq:RPI}
\end{equation}
where $\Omega_L \in \R^{m \times (p+\nu)}$ is a Gaussian random matrix, $p+\nu \ll n$, and $q$ is a small integer (\textit{e.g.,} $q=4$ in \cite{Tepper_2016}). 
This was further extended in \cite{Yahaya_2018} where the authors used RSIs, \textit{i.e.}, a round-off-error stable alternative to RPIs \cite{Halko_2011}.
Lastly, the authors in \cite{Sharan_2019} assumed 
to 
only observe 
$X_L$. 
$F$ and $G_L$ could then be estimated from $X_L$. Assuming the columns of $G$ 
to be sparse w.r.t a known dictionary, they could then be estimated from $G_L$. 

\begin{algorithm}
\caption{Compressed NMF strategy.}\label{algo:compressed_nmf}
\begin{algorithmic}
\REQUIRE initial and compression matrices $G$, $F$, $L$, and $R$.
\STATE Define $ X_L \triangleq L \cdot X$ and $X_R \triangleq X \cdot R$
\REPEAT
\STATE Define $F_R \triangleq F \cdot R  $
\STATE Solve (\ref{eq:NMF_update1}) by resp. replacing $X$ and $F$ by $X_R$ and $F_R$
\STATE Define $G_L \triangleq L \cdot G$
\STATE Solve (\ref{eq:NMF_update2}) by resp. replacing $X$ and $G$ by $X_L$ and $G_L$
\UNTIL{a stopping criterion}
\end{algorithmic}
\end{algorithm}

At this stage, it should be noticed that computing random projections is costly. Indeed, deriving $X_L$ in Algorithm~\ref{algo:compressed_nmf} requires $nm(p+\nu)$ operations. Even worse, computing $L$ in Eq.~(\ref{eq:RPI}) requires---using the Householder QR decomposition---$2 q (p+\nu) n  m + 2 n (p+\nu)^2 - 2/3 (p +\nu)^3$ operations. As a consequence, the authors in \cite{Kapralov_2016} then proposed a cheaper strategy than GC---named CountGauss---which combines the ideas of both the CountSketch method \cite{Clarkson_2013} and GC. The former consists of generating a matrix $S$ with only one nonzero entry per row, whose value is either $+1$ or $-1$ with equal probability. The product $S \cdot X$ then provides a sketch of $X$ 
which is very cheap to compute. However, it requires more samples than GC to reach the same approximation accuracy. Combining CountSketch with Gaussian projection thus leverages the CountSketch drawback while still being faster to compute than a standard Gaussian projection. In that case, applied to NMF, the matrix $L$ reads $L = \Omega_L \cdot S$ where $\Omega_L$ and $S$ have dimensions of size $(p+\nu) \times (p+\mu)$ and $(p+\mu) \times n$, respectively, with $\mu > \nu$ and $(p + \mu) \leq n$. 
Another faster way to compute GC consists of using a dedicated hardware, \textit{e.g.}, Optical Process Unit (OPU) \cite{Saade_2016}. OPUs optically perform random projections, so that they can process very large matrices in a very short time. 
Still, all these alternatives provide a similar performance to GC and should thus be less accurate than RPIs/RSIs when applied to NMF. 

\section{Gaussian Compression Stream}
\label{sec:second-section}
\paragraph{Principle} 
We now introduce our proposed GCS concept. Let us first recall the JLL which states that \cite{Johnson_1984} given $0 < \varepsilon < 1$, a set $X$ of $n$ points in $\mathbb{R}^m$, and a number $k > 8 \log (n)/\varepsilon^2$, there is a linear map $f: \mathbb{R}^m \rightarrow \mathbb{R}^k$ such that $\forall u,v \in X$, $(1-\varepsilon)\|u-v\|^2 \leq \|f(u) - f(v)\|^2 \leq (1+\varepsilon)\|u-v\|^2$. Interestingly, the dimension $k$ of the low-dimensional space only depends on the number $n$ of points in the original high dimensional space and on a distortion parameter $\varepsilon$. Applied to NMF, the linear mapping $f$ is a compression matrix, \textit{i.e.}, $L$ or $R$. In \cite{Tepper_2016}, the authors chose $k \triangleq p +\nu$ where $\nu$ was set to a small value, \textit{i.e.}, $\nu = 10$. This led to a poor NMF performance. However, the JLL implies that by increasing $k$ (or $\nu$), we can reduce the distortion parameter $\varepsilon$ as we less compress the data. However, this implies a reduced speed up of the computations. 

Our proposed strategy thus reads as follows. We assume that $\nu$ is extremely large (or even infinite), so that $L$ and $R$---which are draw according to a scaled Gaussian distribution---cannot fit in memory. We thus assume these matrices to be observed in a streaming fashion, \textit{i.e.}, during the $i$-th NMF iteration, we only observe two $(p+ \nu_i) \times n$ and $m \times (p + \nu_i)$ submatrices of $L$ and $R$, denoted $L^{(i)}$ and $R^{(i)}$, respectively. As a consequence, at each NMF iteration, the update of $G$ and $F$ is done using different compressed matrices $X_R^{(i)}$ and $X_L^{(i)}$, respectively. In the experiments below, we find our strategy to yield the same accuracy as RSIs or vanilla NMF. Still, the approach is only based on GC so that the faster compression strategies discussed above can be applied. However, if we aim to use CPU only, GCS might not be as computationally efficient as RPIs or RSIs as we must compute $X_R^{(i)}$ and $X_L^{(i)}$ at each iteration. We thus also propose an alternative where $L^{(i)}$ and $R^{(i)}$ are updated every $\text{Max}_{\text{Iter}}$ iterations, as shown in Algorithm~\ref{algo:compressed_nmf_GCS}.

\begin{algorithm}
\caption{Proposed compressed NMF strategy with GCS.}\label{algo:compressed_nmf_GCS}
\begin{algorithmic}
\REQUIRE initial matrices $G$, $F$, $i=0$
\REPEAT
\STATE Update $i=i+1$ and get $L^{(i)}$ and $R^{(i)}$
\STATE Define $X_R^{(i)} \triangleq X \cdot R^{(i)}$ and $ X_L^{(i)} \triangleq L^{(i)} \cdot X$
\FOR{counter $=1$ \textbf{to} $\text{Max}_{\text{Iter}}$}
\STATE Define $F_R^{(i)} \triangleq F \cdot R^{(i)}  $ and $G_L^{(i)} \triangleq L^{(i)} \cdot G$ \STATE Solve (\ref{eq:NMF_update1}) by resp. replacing $X$ and $F$ by $X_R^{(i)}$ and $F_L^{(i)}$
\STATE Solve (\ref{eq:NMF_update2}) by resp. replacing $X$ and $G$ by $X_L^{(i)}$ and $G_L^{(i)}$
\ENDFOR
\UNTIL{a stopping criterion}
\end{algorithmic}
\end{algorithm}


\begin{figure}[ht]
~\hfill
\resizebox{.3\textwidth}{!}{
   \begin{tikzpicture}
        \begin{customlegend}[legend columns=3,legend style={align=left,draw=none,column sep=2ex},legend entries={{$\text{Max}_{\text{Iter}}=1$},{$\text{Max}_{\text{Iter}}=2$},{$\text{Max}_{\text{Iter}}=5$},{$\text{Max}_{\text{Iter}}=10$},{$\text{Max}_{\text{Iter}}=25$},{$\text{Max}_{\text{Iter}}=+\infty$ (GC)}}]
        \addlegendimage{red, solid,line width=2pt}
        \addlegendimage{cyan, solid,line width=2pt}
        \addlegendimage{green, solid,line width=2pt}
        \addlegendimage{blue, solid,line width=2pt}
        \addlegendimage{black, solid,line width=2pt}
        \addlegendimage{brown, solid,line width=2pt}
        \end{customlegend}
     \end{tikzpicture}
}\hfill ~\\
 	\resizebox{0.2\textwidth}{!}{\begin{tikzpicture}
 		\begin{semilogyaxis}[
 		ticklabel style = {font=\Large},
 		ytick={1,1.e-2,1.e-4,1.e-6},
 		ymax=1,
 		ymin=1.e-6,
 		xmin=1,
 		xmax=101,
 		xlabel=\large{Iterations},
 		ylabel=\large{RRE},
 		title={GCS-AS-NMF perf. with $\nu_i=10$} 
 		]
  	    \addplot[red,line width=2pt] table[x expr=\coordindex,y=aS_GC_RE, col sep=space]{plots/Dat/r_5_nu_10_pass.dat};
 		\addplot[cyan,line width=2pt] table[x expr=\coordindex,y=aS_GC_RE, col sep=space]{plots/Dat/Omega_2_nu_10_pass.dat};
 		\addplot[green,line width=2pt] table[x expr=\coordindex,y=aS_GC_RE, col sep=space]{plots/Dat/Omega_5_nu_10_pass.dat}; 
 		\addplot[blue,line width=2pt] table[x expr=\coordindex,y=aS_GC_RE, col sep=space]{plots/Dat/Omega_10_nu_10_pass.dat}; 
 		\addplot[black,line width=2pt] table[x expr=\coordindex,y=aS_GC_RE, col sep=space]{plots/Dat/Omega_25_nu_10_pass.dat}; 
 		\addplot[brown,line width=2pt] table[x expr=\coordindex,y=aS_GC, col sep=space]{plots/Dat/Omega_2_nu_10_pass.dat};
 		\end{semilogyaxis}
 		\end{tikzpicture}
 	}\hfill 
 	\resizebox{0.2\textwidth}{!}{\begin{tikzpicture}
 		\begin{semilogyaxis}[
 		ticklabel style = {font=\Large},
 		ytick={1,1.e-2,1.e-4,1.e-6},
 		ymax=1,
 		ymin=1.e-6,
 		xmin=1,
 		xmax=101,
 		xlabel=\large{Iterations},
 		ylabel=\large{RRE},
 		title={GCS-AS-NMF perf. with $\nu_i=50$} 
 		]
  	    \addplot[red,line width=2pt] table[x expr=\coordindex,y=aS_GC_RE, col sep=space]{plots/Dat/r_5_nu_50_pass.dat};
 		\addplot[cyan,line width=2pt] table[x expr=\coordindex,y=aS_GC_RE, col sep=space]{plots/Dat/Omega_2_nu_50_pass.dat};
 		\addplot[green,line width=2pt] table[x expr=\coordindex,y=aS_GC_RE, col sep=space]{plots/Dat/Omega_5_nu_50_pass.dat}; 
 		\addplot[blue,line width=2pt] table[x expr=\coordindex,y=aS_GC_RE, col sep=space]{plots/Dat/Omega_10_nu_50_pass.dat}; 
 		\addplot[black,line width=2pt] table[x expr=\coordindex,y=aS_GC_RE, col sep=space]{plots/Dat/Omega_25_nu_50_pass.dat}; 
 		\addplot[brown,line width=2pt] table[x expr=\coordindex,y=aS_GC, col sep=space]{plots/Dat/Omega_2_nu_50_pass.dat};
 		\end{semilogyaxis}
 		
 		\end{tikzpicture}
 	}
 	
 	\resizebox{0.2\textwidth}{!}{\begin{tikzpicture}
 		\begin{semilogyaxis}[
 		ticklabel style = {font=\Large},
 		ytick={1,1.e-2,1.e-4,1.e-6},
 		ymax=1,
 		ymin=1.e-6,
 		xmin=1,
 		xmax=101,
 		xlabel=\large{Iterations},
 		ylabel=\large{RRE},
 		title={GCS-NeNMF perf. with $\nu_i=10$} 
 		]
  	    \addplot[red,line width=2pt] table[x expr=\coordindex,y=neNMF_GC_RE, col sep=space]{plots/Dat/r_5_nu_10_pass.dat};
 		\addplot[cyan,line width=2pt] table[x expr=\coordindex,y=neNMF_GC_RE, col sep=space]{plots/Dat/Omega_2_nu_10_pass.dat};
 		\addplot[green,line width=2pt] table[x expr=\coordindex,y=neNMF_GC_RE, col sep=space]{plots/Dat/Omega_5_nu_10_pass.dat}; 
 		\addplot[blue,line width=2pt] table[x expr=\coordindex,y=neNMF_GC_RE, col sep=space]{plots/Dat/Omega_10_nu_10_pass.dat}; 
 		\addplot[black,line width=2pt] table[x expr=\coordindex,y=neNMF_GC_RE, col sep=space]{plots/Dat/Omega_25_nu_10_pass.dat}; 
 		\addplot[brown,line width=2pt] table[x expr=\coordindex,y=neNMF_GC, col sep=space]{plots/Dat/Omega_2_nu_10_pass.dat};
 		\end{semilogyaxis}
 		\end{tikzpicture}
 	}\hfill
 	\resizebox{0.2\textwidth}{!}{\begin{tikzpicture}
 		\begin{semilogyaxis}[
 		ticklabel style = {font=\Large},
 		ytick={1,1.e-2,1.e-4,1.e-6},
 		ymax=1,
 		ymin=1.e-6,
 		xmin=1,
 		xmax=101,
 		xlabel=\large{Iterations},
 		ylabel=\large{RRE},
 		title={GCS-NeNMF perf. with $\nu_i=50$}
 		]
  	    \addplot[red,line width=2pt] table[x expr=\coordindex,y=neNMF_GC_RE, col sep=space]{plots/Dat/r_5_nu_50_pass.dat};
 		\addplot[cyan,line width=2pt] table[x expr=\coordindex,y=neNMF_GC_RE, col sep=space]{plots/Dat/Omega_2_nu_50_pass.dat};
 		\addplot[green,line width=2pt] table[x expr=\coordindex,y=neNMF_GC_RE, col sep=space]{plots/Dat/Omega_5_nu_50_pass.dat}; 
 		\addplot[blue,line width=2pt] table[x expr=\coordindex,y=neNMF_GC_RE, col sep=space]{plots/Dat/Omega_10_nu_50_pass.dat}; 
 		\addplot[black,line width=2pt] table[x expr=\coordindex,y=neNMF_GC_RE, col sep=space]{plots/Dat/Omega_25_nu_50_pass.dat}; 
 		\addplot[brown,line width=2pt] table[x expr=\coordindex,y=neNMF_GC, col sep=space]{plots/Dat/Omega_2_nu_50_pass.dat};
 		\end{semilogyaxis}
 		\end{tikzpicture}
 	}

 \caption{NMF performance for different parameters of the GCS strategy.}\label{fig:RRE1}
\end{figure}
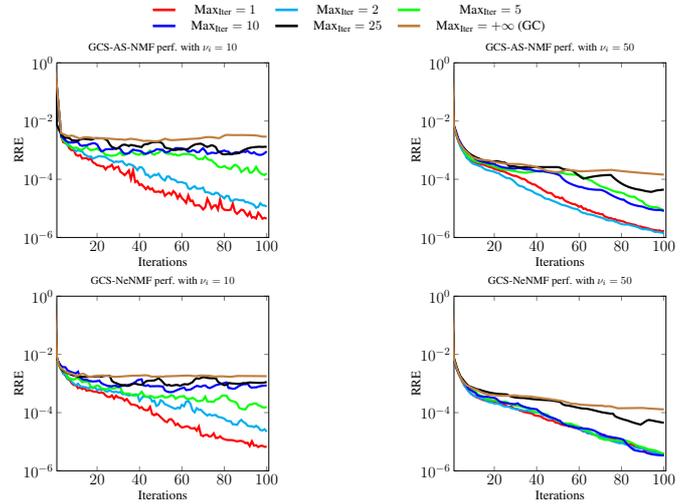

\paragraph{Experiments} 
We now investigate the performance of our proposed method. For that purpose, we consider 15 simulations where we draw random nonnegative matrices $G$ and $F$ such that $n=m=10000$ and $p=5$. As a consequence, their product $X$ is a $10000 \times 10000$ rank-5 matrix. To assess the performance of the proposed method, we consider two different NMF solvers, \textit{i.e.}, Active Set (AS-NMF) \cite{Kim_2008} and Nesterov gradient (NeNMF) \cite{Guan_2012} and three different compression strategies, \textit{i.e.}, RSIs, GC and GCS, that we compare to the performance reached by the vanilla strategy. The performance criterion used in this paper is a Relative Reconstruction Error (RRE), 
defined as $\text{RRE} \triangleq \normF{X-G\cdot F}^2/\normF{X}^2$.
In each simulation, we consider the same random initialization and we compute the RRE performance for each tested method at each NMF iteration. Figure~\ref{fig:RRE1} shows the median RREs achieved by AS-NMF and NeNMF combined with GCS with respect to $\text{Max}_{\text{Iter}}$ and $\nu_i$. We notice that (i) GCS always outperforms GC 
, (ii) the RREs are not always decreasing along iterations when $\nu_i=10$, and (iii) the best performance is achieved when $\text{Max}_{\text{Iter}}=1$ or 2. 

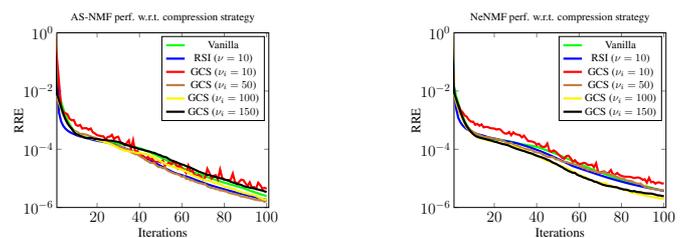
\begin{figure}[ht]
  	\resizebox{0.2\textwidth}{!}{\begin{tikzpicture}
 		 \begin{semilogyaxis}[
 		ticklabel style = {font=\Large},
 		ytick={1,1.e-2,1.e-4,1.e-6},
 		ymax=1,
 		ymin=1.e-6,
 		xmin=1,
 		xmax=101,
 		xlabel=\large{Iterations},
 		ylabel=\large{RRE},
 		title={AS-NMF perf. w.r.t. compression strategy}
 		]
 	 \addplot[green,line width=2pt] table[x expr=\coordindex, y=aS_V, col sep=space]{plots/Dat/r_5_nu_10_pass.dat};
 	\addplot[blue,line width=2pt] table[x expr=\coordindex,y=aS_RSI, col sep=space]{plots/Dat/r_5_nu_10_pass.dat};
 	\addplot[red,line width=2pt] table[x expr=\coordindex,y=aS_GC_RE, col sep=space]{plots/Dat/r_5_nu_10_pass.dat};
 	\addplot[brown,line width=2pt] table[x expr=\coordindex,y=aS_GC_RE, col sep=space]{plots/Dat/r_5_nu_50_pass.dat};
 	\addplot[yellow,line width=2pt] table[x expr=\coordindex,y=aS_GC_RE, col sep=space]{plots/Dat/r_5_nu_100_pass.dat};
 	\addplot[black,line width=2pt] table[x expr=\coordindex,y=aS_GC_RE, col sep=space]{plots/Dat/r_5_nu_150_pass.dat};
 	\legend{{Vanilla},{RSI ($\nu=10$)},{GCS ($\nu_i=10$)},{GCS ($\nu_i=50$)},{GCS ($\nu_i=100$)},{GCS ($\nu_i=150$)}}
 		\end{semilogyaxis}
 		
 		\end{tikzpicture}
 	} \hfill
  	\resizebox{0.2\textwidth}{!}{\begin{tikzpicture}
 		\begin{semilogyaxis}[
 		ticklabel style = {font=\Large},
 		ytick={1,1.e-2,1.e-4,1.e-6},
 		ymax=1,
 		ymin=1.e-6,
 		xmin=1,
 		xmax=101,
 		xlabel=\large{Iterations},
 		ylabel=\large{RRE},
 		title={NeNMF perf. w.r.t. compression strategy}
 		]
 		\addplot[green,line width=2pt] table[x expr=\coordindex, y=neNMF_V, col sep=space]{plots/Dat/r_5_nu_10_pass.dat};
 		\addplot[blue,line width=2pt] table[x expr=\coordindex,y=neNMF_RSI, col sep=space]{plots/Dat/r_5_nu_10_pass.dat};
 		\addplot[red,line width=2pt] table[x expr=\coordindex,y=neNMF_GC_RE, col sep=space]{plots/Dat/r_5_nu_10_pass.dat};
 		 \addplot[brown,line width=2pt] table[x expr=\coordindex,y=neNMF_GC_RE, col sep=space]{plots/Dat/r_5_nu_50_pass.dat};
 	    \addplot[yellow,line width=2pt] table[x expr=\coordindex,y=neNMF_GC_RE, col sep=space]{plots/Dat/r_5_nu_100_pass.dat};
 	    \addplot[black,line width=2pt] table[x expr=\coordindex,y=neNMF_GC_RE, col sep=space]{plots/Dat/r_5_nu_150_pass.dat};
 	\legend{{Vanilla},{RSI ($\nu=10$)},{GCS ($\nu_i=10$)},{GCS ($\nu_i=50$)},{GCS ($\nu_i=100$)},{GCS ($\nu_i=150$)}}
 		\end{semilogyaxis}
 		
 		\end{tikzpicture}
 	} 
 	\caption{NMF performance with respect to compression techniques.} 
 	\label{fig:RRE2}
 \end{figure}
Figure~\ref{fig:RRE2} shows the evolution of the median RREs with no compression, RSIs, and GCS (with $\text{Max}_{\text{Iter}}=1$ but for different values of $\nu_i$). The latter provides a similar or better enhancement than the former except $\nu_i = 10$.  

\paragraph{Conclusion} \label{sec:conclusion}
We proposed a new randomized compression technique which is based on GC only and which is shown to be accurate on the considered NMF simulations. However, it requires the computation of the compressed matrices every $\text{Max}_{\text{Iter}}$ iterations, which might be prohibitive w.r.t. structured compression if the number of NMF iterations is high. This issue 
might be solved using a dedicated hardware. 
We will investigate it in our future work and we will investigate the use of GCS for mobile sensor calibration \cite{Dorffer_2018,Dorffer_2016b}.
 
\paragraph{Acknowledgements} F. Yahaya gratefully acknowledges the R\'egion Hauts-de-France to partly fund his PhD fellowship. 
Experiments presented in this paper were carried out using the CALCULCO computing platform, supported by SCoSI/ULCO.

\bibliographystyle{IEEEbib}
\bibliography{IEEEabrv,biblio.bib}

\begin{thebibliography}{10}

\bibitem{Halko_2011}
N.~Halko, P.-G. Martinsson, and J.~A Tropp,
\newblock ``Finding structure with randomness: Probabilistic algorithms for
  constructing approximate matrix decompositions,''
\newblock {\em SIAM review}, vol. 53, no. 2, pp. 217--288, 2011.

\bibitem{Drineas_2016}
P.~Drineas and M.~W. Mahoney,
\newblock ``{RandNLA}: randomized numerical linear algebra,''
\newblock {\em Communications of the ACM}, vol. 59, no. 6, pp. 80--90, 2016.

\bibitem{Tepper_2016}
M.~Tepper and G.~Sapiro,
\newblock ``Compressed nonnegative matrix factorization is fast and accurate,''
\newblock {\em {IEEE} Trans. Signal Process.}, vol. 64, no. 9, pp. 2269--2283,
  May 2016.

\bibitem{Erichson_2018}
N.~B. Erichson, A.~Mendible, S.~Wihlborn, and J~N. Kutz,
\newblock ``Randomized nonnegative matrix factorization,''
\newblock {\em Pattern Recognition Letters}, 2018.

\bibitem{Yahaya_2018}
F.~Yahaya, M.~Puigt, G.~Delmaire, and G.~Roussel,
\newblock ``Faster-than-fast {NMF} using random projections and {N}esterov
  iterations,''
\newblock in {\em Proc. iTWIST'18}, 2018.

\bibitem{Yahaya_2019}
F.~Yahaya, M.~Puigt, G.~Delmaire, and G.~Roussel,
\newblock ``How to apply random projections to nonnegative matrix factorization
  with missing entries?,''
\newblock in {\em Proc. EUSICPO'19}, 2019.

\bibitem{Kapralov_2016}
M.~Kapralov, V.~Potluru, and D.~Woodruff,
\newblock ``How to fake multiply by a gaussian matrix,''
\newblock in {\em Proc. ICML'16}, 2016, pp. 2101--2110.

\bibitem{Saade_2016}
A.~Saade, F.~Caltagirone, I.~Carron, L.~Daudet, A.~Dr\'emeau, S.~Gigan, and
  F.~Krzakala,
\newblock ``Random projections through multiple optical scattering:
  Approximating kernels at the speed of light,''
\newblock in {\em Proc. ICASSP'16}, 2016, pp. 6215--6219.

\bibitem{Wang_2013}
Y.~X. Wang and Y.~J. Zhang,
\newblock ``Nonnegative matrix factorization: A comprehensive review,''
\newblock {\em {IEEE} Trans. Knowl. Data Eng.}, vol. 25, no. 6, pp. 1336--1353,
  June 2013.

\bibitem{Gillis_2014}
N.~Gillis,
\newblock ``The why and how of nonnegative matrix factorization,''
\newblock in {\em Regularization, Optimization, Kernels, and Support Vector
  Machines}, pp. 257--291. Chapman and Hall/CRC, 2014.

\bibitem{Liu_2010}
C.~Liu, H.-C. Yang, J.~Fan, L.-W. He, and Y.-M. Wang,
\newblock ``Distributed nonnegative matrix factorization for web-scale dyadic
  data analysis on {MapReduce},''
\newblock in {\em Proc. WWW Conf.'10}, April 2010.

\bibitem{Mairal_2010}
J.~Mairal, F.~Bach, J.~Ponce, and G.~Sapiro,
\newblock ``Online learning for matrix factorization and sparse coding,''
\newblock {\em Journal of Machine Learning Research}, vol. 11, no. Jan, pp.
  19--60, 2010.

\bibitem{Guan_2012}
N.~Guan, D.Vin Tao, Z.~Luo, and B.~Yuan,
\newblock ``{NeNMF}: An optimal gradient method for nonnegative matrix
  factorization,''
\newblock {\em {IEEE} Trans. Signal Process.}, vol. 60, no. 6, pp. 2882--2898,
  2012.

\bibitem{Zhou_2012}
G.~Zhou, A.~Cichocki, and S.~Xie,
\newblock ``Fast nonnegative matrix/tensor factorization based on low-rank
  approximation,''
\newblock {\em {IEEE} Trans. Signal Process.}, vol. 60, no. 6, pp. 2928--2940,
  June 2012.

\bibitem{Wang_2010}
F.~Wang and P.~Li,
\newblock ``Efficient nonnegative matrix factorization with random
  projections,''
\newblock in {\em Proc. SIAM ICDM'10}. SIAM, 2010, pp. 281--292.

\bibitem{Johnson_1984}
W.~B. Johnson and J.~Lindenstrauss,
\newblock ``Extensions of {Lipschitz} mappings into a {Hilbert} space,''
\newblock {\em Contemporary mathematics}, vol. 26, no. 189-206, pp. 1, 1984.

\bibitem{Sharan_2019}
V.~Sharan, K.~S. Tai, P.~Bailis, and G.~Valiant,
\newblock ``Compressed factorization: Fast and accurate low-rank factorization
  of compressively-sensed data,''
\newblock in {\em Proc. ICML'19}, 2019, pp. 5690--5700.

\bibitem{Clarkson_2013}
K.~L. Clarkson and D.~P. Woodruff,
\newblock ``Low rank approximation and regression in input sparsity time,''
\newblock in {\em Proc. ACM STOC'13}, New York, NY, USA, 2013, pp. 81--90,
  Association for Computing Machinery.

\bibitem{Kim_2008}
H.~Kim and H.~Park,
\newblock ``Nonnegative matrix factorization based on alternating nonnegativity
  constrained least squares and active set method,''
\newblock {\em SIAM Journal on Matrix Analysis and Applications}, vol. 30, no.
  2, pp. 713--730, 2008.

\bibitem{Dorffer_2018}
C.~Dorffer, M.~Puigt, G.~Delmaire, and G.~Roussel,
\newblock ``Informed nonnegative matrix factorization methods for mobile sensor
  network calibration,''
\newblock {\em {IEEE} Trans. Signal Inf. Process. Netw.}, vol. 4, no. 4, pp.
  667--682, Dec 2018.

\bibitem{Dorffer_2016b}
C.~Dorffer, M.~Puigt, G.~Delmaire, and G.~Roussel,
\newblock ``Nonlinear mobile sensor calibration using informed semi-nonnegative
  matrix factorization with a {V}andermonde factor,''
\newblock in {\em Proc. SAM'16}, 2016.

\end{thebibliography}

\end{document}